\documentclass[aps, prd, twocolumn, superscriptaddress]{revtex4-1}
\usepackage{graphicx}
\usepackage{amssymb}
\usepackage{amsmath}
\usepackage{mathtools}
\usepackage{lmodern}
\usepackage{booktabs}
\usepackage{bbold}
\usepackage{bm}
\usepackage{bbm}
\usepackage[colorlinks=True,linkcolor=blue,anchorcolor=blue,citecolor=blue,CJKbookmarks=true]{hyperref}

\usepackage{ulem}

\begin{document}
\title{Study of spin states in vacuum pair production via the Dirac-Heisenberg-Wigner formalism}

\author{R. Z. Jiang}
\affiliation{State Key Laboratory for Tunnel Engineering, China University of Mining and Technology, Beijing 100083, China}

\author{Z. L. Li}
\email{Corresponding author. zlli@cumtb.edu.cn}
\affiliation{State Key Laboratory for Tunnel Engineering, China University of Mining and Technology, Beijing 100083, China}
\affiliation{School of Science, China University of Mining and Technology, Beijing 100083, China}

\author{Y. J. Li}
\email{Corresponding author. lyj@aphy.iphy.ac.cn}
\affiliation{State Key Laboratory for Tunnel Engineering, China University of Mining and Technology, Beijing 100083, China}
\affiliation{School of Science, China University of Mining and Technology, Beijing 100083, China}

\date{\today}

\begin{abstract}
A general spin-resolved momentum distribution of electron-positron pairs produced in strong external fields is derived by combining the covariant spin projection operator and the Dirac-Heisenberg-Wigner (DHW) formalism.
The result shows that the spin-resolved and helicity-resolved momentum distributions given in previous literature are actually two special cases of it.
For any spin-direction unit vector, numerical investigations demonstrate that when the $z$-component of the unit vector vanishes, the number density of produced spin-up and spin-down particles is equal, while their momentum distributions have some asymmetry.
For a nonzero $z$-component of the unit vector, there is a difference of $1-3$ orders of magnitude in the number density of spin-up and spin-down particles induced by angular momentum transfer in multiphoton absorption.
Moreover, as the electric field strength increases and/or the field frequency decreases, the asymmetry between the spin-up and spin-down particle number density decreases rapidly.
These results offer an approach to study general spin states in vacuum pair production, and enhance our understanding of angular momentum transfer from fields to matter in extreme environments.
\end{abstract}

\maketitle

\section{INTRODUCTION}
\label{sec:one}
In the presence of a strong external electromagnetic field, the quantum electrodynamics (QED) vacuum becomes unstable and decays into electron-positron pairs \cite{Sauter1931,Heisenberg1936,Schwinger1951,DiPiazza2011,Xie2017}.
Since Dirac proposed the relativistic wave equation and predicted the existence of the positron \cite{Dirac1928}, this topic has always been one of the most fascinating theoretical predictions in QED.
Using the proper-time method, Schwinger calculated the electron-positron pair production rate in a strong constant field, which is $\sim\exp(-\pi E_{\mathrm{cr}}/E_0)$ \cite{Schwinger1951}, where $E_0$ is the constant field strength, $E_{\mathrm{cr}}=m^2c^3/(e\hbar)\sim10^{16}\mathrm{V}/\mathrm{cm}$ is the so-called Schwinger critical field strength, and $m$ and $e$ are the magnitudes of the mass and charge of the electrons, respectively.
Because of the exponential suppression of the production rate and the fact that the electric field strength presently achievable in the laboratory is much lower than the critical field strength $E_{\mathrm{cr}}$, the phenomenon has not been verified directly in experiments so far.
Nevertheless, the development of high-intensity laser technology \cite{ELI,XCELS} provides new possibilities for observing electron-positron pair production in vacuum environments in the future.

In terms of theoretical research, significant progress has been made on the topic of vacuum pair production.
Researchers have not only developed a variety of nonperturbative computational methods, but also revealed important physical phenomena, such as the dynamically assisted Schwinger mechanism \cite{Schutzhold2008,Torgrimsson2017,Torgrimsson2018,Taya2020,Li2021L}, the effective mass signatures \cite{Kohlfurst2014}, the spiral structures of the momentum distribution \cite{Li2017,Li2019,Hu2023}, Ramsey interference \cite{Akkermans2012,Li2014}, and so on.
The widely used research methods include computational quantum field theory (CQFT) \cite{su2019,Zhou2021,jiang2023}, the worldline instanton technique \cite{Dunne2005,Dunne2006,Dumlu2011,Rajeev2021,Orkash2022}, Wentzel-Kramers-Brillouin (WKB) approximation \cite{Dunne2011,Schutzhold2019}, Dirac-Heisenberg-Wigner (DHW) formalism \cite{Bialynicki1991,Kohlfurst2018,Kohlfurst2020,Yu2023}, and so on.
The relationship or equivalence between different methods has also been explored \cite{Dumlu2009Q,Hebenstreit2010,Strobel2014,Li2021A}, which assists in understanding the theoretical results from different perspectives.
Recently, Aleksandrov $et\,\,al.$ \cite{Aleksandrov2024k} corrected the quantum kinetic equations (QKEs) for studying electron-positron pair production in spatially homogeneous and time-dependent electric fields with multiple components and proved that they are equivalent to DHW formalism.
Moreover, spin-resolved and helicity-resolved momentum distribution functions are also discussed.

In fact, the phenomena related to spin and/or helicity in electron-positron pair production from vacuum have also attracted widespread attention \cite{Strobel2015, Blinne2016c, Li2019B, Kohlfurst2019s, Hu2024, Aleksandrov2024A,Majczak2024,Amat2025,Chen2025xib}, leading to a series of important developments.
In Ref. \cite{Strobel2015}, Strobel and Xue studied the pair production rate for rotating electric fields with the semiclassical WKB approach and found an asymmetry in the spin states.
Blinne and Strobel \cite{Blinne2016c} recovered this asymmetry using the projection operator in the framework of the DHW formalism.
They proposed that the quantum state corresponding to the projection operator of a given physical quantity can be extracted from the Wigner function.
Based on this, one can obtain the momentum distribution functions with additional observables.
Li $et\,\,al.$ \cite{Li2019B} revealed that for a multicycle elliptically polarized field, the spin asymmetry roughly increases with the Keldysh adiabatic parameter \cite{Keldysh1965}.
By studying electron-positron pairs produced by two circularly polarized fields with different frequencies and a time delay, Hu $et\,\,al.$ \cite{Hu2024} found that a spiral structure exists in both the spin-up and spin-down momentum distributions, and the spin asymmetry degree can be significantly altered by adjusting the time delay between these two fields.
In addition, Aleksandrov and Kudlis \cite{Aleksandrov2024A} analyzed helicity-resolved momentum distributions of produced particles by left-handed rotating electric fields, and found that left-handed electrons (negative helicity) always prefer to propagate along the negative $z$-axis, while right-handed electrons (positive helicity) always prefer to propagate along the positive $z$-axis.

This paper focuses on the spin states of electron-positron pairs produced in a circularly polarized electric field.
There are three main contents: First, a general expression for the momentum distribution of produced particles with different spin states is derived by combining the covariant spin projection operator with the DHW formalism.
When the spin-direction unit vector is along the direction of particle momentum, we can obtain the same helicity-resolved momentum distribution as that in Ref. \cite{Aleksandrov2024A}.
When the unit vector is taken along the $z$-axis, the spin-resolved momentum distribution is obtained, and it is identical to that in Ref. \cite{Aleksandrov2024k}.
It is also clarified that the spin-resolved momentum distribution derived from the semiclassical WKB approximation \cite{Strobel2015} and the DHW formalism \cite{Blinne2016c,Hu2024,Chen2025xib} do not correspond to genuine pure spin states when the momentum of created particles along the field direction is not zero.
The second is the study of the spin-resolved momentum distribution with any spin-direction unit vector.
Thirdly, the effect of electric field strength and frequency on the spin asymmetry is examined.

The structure of this paper is as follows:
In Sec. \ref{sec:two}, to make this work self-contained, we briefly review the DHW formalism and QKEs.
In Sec. \ref{sec:three}, the theoretical formulas describing spin- and helicity-resolved momentum distribution functions are derived by combining the covariant spin projection operator with the DHW formalism.
In Sec. \ref{sec:four}, we investigate the spin-resolved momentum distribution and number density of particles produced in a circularly polarized electric field.
Finally, the summary is given in Sec. \ref{sec:five}.
In this paper, natural units $(\hbar=c=1)$ are used.

\section{DIRAC-HEISENBERG-WIGNER FORMALISM AND QUANTUM KINETIC EQUATIONS}
\label{sec:two}
The study in this paper is mainly based on the DHW formalism, which has been widely applied to investigate electron-positron pair production from vacuum by strong background fields.
The DHW formalism can not only provide information about the time evolution of the single-particle momentum distribution function conveniently \cite{Blinne2014p, Diez2023}, but also offer additional observables by combining with projection operators \cite{Blinne2016c, Blinne2016d}.
Since our study also utilized QKEs to analyze the results, we provide a brief review of the DHW formalism and QKEs in this section.

\subsection{Dirac-Heisenberg-Wigner (DHW) formalism}\label{subsec:A}
We start from the gauge-invariant density operator consisting of two Dirac field operators in the Heisenberg picture,
\begin{eqnarray}\label{eq:GCDO}
\hat{\mathcal{C}}_{\alpha \beta}\left( \mathbf{x},\mathbf{y},t \right)&=&\exp \left( -ie\int_{-1/2}^{1/2}{\mathbf{A}\left( \mathbf{x}+\xi \mathbf{y}, t \right) \cdot \mathbf{y}d\xi} \right)\nonumber\\
&&\times \left[ \hat{\Psi}_{\alpha}\left( \mathbf{x}+\frac{\mathbf{y}}{2}, t \right) , \hat{\bar{\Psi}}_{\beta}\left( \mathbf{x}-\frac{\mathbf{y}}{2}, t \right) \right],
\end{eqnarray}
where $\mathbf{A}$ is the vector potential, and $\mathbf{x}$ and $\mathbf{y}$ denote the center of mass and the relative coordinates, respectively.
Note that the temporal gauge and the mean-field approximation to the electromagnetic field are used.

The key quantity in the DHW formalism is the Wigner operator, which is defined as the Fourier transform of the Eq. (\ref{eq:GCDO}) with respect to $\mathbf{y}$. Its vacuum expectation is known as the Wigner function,
\begin{eqnarray}\label{eq:VEWF}
\mathcal{W} \left( \mathbf{x},\mathbf{p},t \right) =-\frac{1}{2}\int{d^3ye^{-i\mathbf{p}\cdot \mathbf{y}}}\left< 0\left| \hat{\mathcal{C}}_{\alpha \beta}\left( \mathbf{x},\mathbf{y},t \right) \right|0 \right>.
\end{eqnarray}
Employing a set of complete bases $\left\{ \mathbbm{1} ,\gamma _5,\gamma ^{\mu},\gamma ^{\mu}\gamma _5,\sigma ^{\mu \nu} \right\} $ of the Dirac algebra, the Wigner function can be decomposed into $16$ real functions:
\begin{eqnarray}\label{eq:DVEWF}
\mathcal{W} \left( \mathbf{x},\mathbf{p},t \right) =\frac{1}{4}\left( \mathbbm{1} \mathbbm{s} +i\gamma _5\mathbbm{p} +\gamma ^{\mu}\mathbbm{v} _{\mu}+\gamma ^{\mu}\gamma _5\mathbbm{a} _{\mu}+\sigma ^{\mu \nu}\mathbbm{t} _{\mu \nu} \right),
\end{eqnarray}
where $\gamma_5=i\gamma^0\gamma^1\gamma^2\gamma^3$ and $\sigma^{\mu\nu}=\frac{i}{2}[\gamma ^{\mu}, \gamma^{\nu}]$. $\mathbbm{s}$, $\mathbbm{p}$, $\mathbbm{v}_{\mu}$, $\mathbbm{a}_{\mu}$, and $\mathbbm{t}_{\mu \nu}$ denote scalar, pseudoscalar, vector, axis vector, and tensor, respectively, which are also known as DHW functions.
To make the description convenient, we can decompose the $4$-vector and tensor into $\mathbbm{v}^{\mu}=(\mathbb{v}^0, \mathbf{v})$, $\mathbbm{a}^{\mu}=(\mathbbm{a}^0, \mathbf{a})$, $(\mathbf{t}_1)^i=\mathbbm{t}^{i0}-\mathbbm{t}^{0i}$, and $\mathbf{t}_2=\epsilon^{ijk}\mathbbm{t}_{ij}\mathbf{e}_k$.
Here, $\mathbf{e}_k$ is the unit vector.
The partial differential equation system for the $16$ Wigner function components described above can be obtained by taking Eq. (\ref{eq:DVEWF}) into the equation of motion satisfied by the Wigner function, more details can be found in Refs. \cite{Hebenstreit2016phd, Kohlfurst2015phd}.
When the system is in the initial vacuum state, the non-zero terms of the Wigner function are
\begin{eqnarray}\label{eq:INZDHW}
\mathbbm{s}_{\mathrm{vac}}=-\frac{2m}{\omega \left( \mathbf{p} \right)},\quad \mathbf{v}_{\mathrm{vac}}=-\frac{2\mathbf{p}}{\omega \left( \mathbf{p} \right)}.
\end{eqnarray}
Here $\omega =\sqrt{m^2+\mathbf{p}^2}$ denotes the energy of the particle, and $\mathbf{p}$ represents the kinetic momentum of the particle.

In the following, we focus on the case of a spatially homogeneous and time-dependent electric field.
Under this condition, $\mathbbm{p}$, $\mathbbm{v}^0$, $\mathbbm{a}^0$, and $\mathbf{t}_2$ vanish.
Using the method of characteristics and replacing the kinetic momentum $\mathbf{p}=\mathbf{q}-e\mathbf{A}(t)$ by the canonical momentum $\mathbf{q}$, the system of partial differential equations with $16$ DHW functions can be transformed into an ordinary differential equation system consisting of the remaining $10$ nonzero DHW functions $W( \mathbf{q},t)=(\mathbbm{s},\mathbf{v},\mathbf{a},
\mathbf{t}_1)^{\intercal}(\mathbf{q},t)$,
\begin{eqnarray}\label{eq:NDHWODES}
\dot{W}\left( \mathbf{q},t \right) =\mathcal{M} \left( \mathbf{q},t \right) W\left( \mathbf{q},t \right).
\end{eqnarray}
Here the dot above the letter denotes the derivative with respect to time and $\mathcal{M}$ is a $10\times10$ matrix:
\begin{eqnarray}\label{eq:MATRIXM}
\mathcal{M} \left( \mathbf{q},t \right) =\left( \begin{matrix}
	0&		0&		0&		2\mathbf{p}^{\intercal}\\
	0&		0&		-2\mathbf{p}\times&		-2m\\
	0&		-2\mathbf{p}\times&		0&		0\\
	-2\mathbf{p}&		2m&		0&		0\\
\end{matrix} \right).
\end{eqnarray}

The single-particle momentum distribution function $f(\mathbf{q},t)$ is defined as
\begin{eqnarray}\label{eq:DFSPMD}
f\left( \mathbf{q},t \right) =\frac{1}{2\omega \left( \mathbf{q},t \right)}\left[ \varepsilon \left( \mathbf{q},t \right) -\varepsilon _{\mathrm{vac}}\left( \mathbf{q},t \right) \right],
\end{eqnarray}
where $\varepsilon=m\mathbbm{s}+\mathbf{p}\cdot \mathbf{v}$ represents the phase space energy density.
For numerical computation, a vector auxiliary function is introduced,
\begin{eqnarray}\label{eq:VAF}
\boldsymbol{v} \left( \mathbf{q},t \right) =-\mathbf{v}\left( \mathbf{q},t \right) +\left[ 1-f\left( \mathbf{q},t \right) \right] \mathbf{v}_{\mathrm{vac}}\left( \mathbf{q},t \right).
\end{eqnarray}
Then the system of Eq. (\ref{eq:NDHWODES}) has the following form:
\begin{eqnarray}\label{eq:FVAT}
\begin{array}{l}
\dot{f}=\displaystyle{\frac{e\mathbf{E}\cdot \boldsymbol{v}}{2\omega}},
\\
\\
\dot{\boldsymbol{v}}=\displaystyle{\frac{2\left[ \left( e\mathbf{E}\cdot \mathbf{p} \right) \mathbf{p}-\omega ^2e\mathbf{E} \right]}{\omega ^3}\left( f-1 \right) -\frac{\left( e\mathbf{E}\cdot \boldsymbol{v} \right) \mathbf{p}}{\omega ^2}}
\\
\\
\quad \quad     \displaystyle{-2\mathbf{p}\times \mathbf{a}-2m\mathbf{t}_1},
\\
\\
\dot{\mathbf{a}}=\displaystyle{-2\mathbf{p}\times \boldsymbol{v}} ,
\\
\\
\dot{\mathbf{t}}_1=\displaystyle{\frac{2}{m}\left[ m^2\boldsymbol{v} +\left( \mathbf{p}\cdot \boldsymbol{v} \right) \mathbf{p} \right]}.
\end{array}
\end{eqnarray}
Here $\mathbf{E}=-\dot{\mathbf{A}}(t)$ is a spatially homogeneous and time-dependent electric field.
The single-particle momentum distribution function $f(\mathbf{q},t)$ is then obtained by solving the system of Eq. (\ref{eq:FVAT}) with the initial conditions $f(\mathbf{q},-\infty)=0$ and $\boldsymbol{v}(\mathbf{q},-\infty)=\mathbf{a}(\mathbf{q},-\infty)=\mathbf{t}_1( \mathbf{q},-\infty)=0$.
In addition, the number density of produced particles can be obtained by integrating $f(\mathbf{q},t)$ over the full momentum space at $t\rightarrow +\infty$, i.e.,
\begin{eqnarray}\label{eq:NDOP}
N=\underset{t\rightarrow +\infty}{\lim}\int{\frac{d^3q}{\left( 2\pi \right)^3}}f\left( \mathbf{q},t \right).
\end{eqnarray}

\subsection{Quantum kinetic equations (QKEs)}\label{subsec:B}
Under a spatially homogeneous and time-dependent electric field, the QKEs consist of an ordinary differential equations system with $10$ components.
The full details of the derivation are given in Ref. \cite{Aleksandrov2024k}.
Here we only give the results:
\begin{eqnarray}\label{eq:QKES}
\begin{array}{l}
\dot{f}=\displaystyle{-4\boldsymbol{\xi }_2\cdot \boldsymbol{\mu }},
\\
\\
\dot{\boldsymbol{\lambda}}=\displaystyle{2\boldsymbol{\xi }_1\times \boldsymbol{\lambda }-2\boldsymbol{\xi }_2\times \boldsymbol{\nu }},
\\
\\
\dot{\boldsymbol{\mu}}=\displaystyle{2\boldsymbol{\xi }_1\times \boldsymbol{\mu }+\boldsymbol{\xi }_2\left( f-1 \right) +2\omega \boldsymbol{\nu }},
\\
\\
\dot{\boldsymbol{\nu}}=\displaystyle{2\boldsymbol{\xi }_1\times \boldsymbol{\nu }-2\boldsymbol{\xi }_2\times \boldsymbol{\lambda }-2\omega \boldsymbol{\mu }}.
\end{array}
\end{eqnarray}
Where $\boldsymbol{\xi }_1$ and $\boldsymbol{\xi }_2$ are two functions orthogonal to each other,
\begin{eqnarray}\label{eq:QKESXI}
\begin{array}{l}
\boldsymbol{\xi }_1=\displaystyle{\frac{e}{2\omega \left( \omega +m \right)}\mathbf{p}\times \mathbf{E}},
\\
\\
\boldsymbol{\xi }_2=\displaystyle{\frac{e}{2\omega ^2\left( \omega +m \right)}\left[ \left( \mathbf{p}\cdot \mathbf{E} \right) \mathbf{p}-\omega \left( \omega +m \right) \mathbf{E} \right]} .
\end{array}
\end{eqnarray}
All components of the QKEs are $0$ at the initial vacuum state.
Note that the single-particle momentum distribution function $f$ here is the same as we defined in the DHW formalism.

\section{SPIN- AND HELICITY-RESOLVED MOMENTUM DISTRIBUTION FUNCTIONS}
\label{sec:three}
Besides the full single-particle momentum distribution function, various additional observable quantum state information can be extracted from the Wigner function.
Applying the projection operator $\mathcal{P}$ to the Wigner function results in a projected single-particle momentum distribution function \cite{Blinne2016c, Blinne2016d},
\begin{eqnarray}\label{eq:PSMDF}
f_{\mathcal{P}}\coloneqq \frac{1}{2\omega}\mathrm{tr}\left[ \mathcal{P} @\left( \mathcal{W} -\mathcal{W} _{\mathrm{vac}} \right) \left( m\mathbbm{1} +\bm{\gamma}\cdot\mathbf{p} \right) \right],
\end{eqnarray}
where
\begin{eqnarray}\label{eq:PROOPE}
\mathcal{P} @\mathcal{W} \coloneqq \frac{1}{2}\left( \mathcal{P} \mathcal{W} +\mathcal{W} \gamma ^0\mathcal{P} ^{\dagger}\gamma ^0 \right).
\end{eqnarray}

If one wants to obtain the spin-resolved single-particle momentum distribution function from the Wigner function, it is necessary to utilize the covariant spin projection operator \cite{Lahiri2005, Greiner2000},
\begin{eqnarray}\label{eq:SPO}
\mathcal{P} _{s}^{\pm}=\frac{1}{2}\left( \mathbbm{1}\pm \gamma _5\gamma ^{\mu}n_{\mu} \right),
\end{eqnarray}
where $+$ and $-$ represent spin up and spin down, respectively.
And the $4$-vector $n^{\mu}=(n^0, \mathbf{n})$ is
\begin{eqnarray}\label{eq:NVECTOR}
n^0=\frac{\mathbf{p}\cdot \mathbf{s}}{m},\quad \mathbf{n}=\mathbf{s}+\frac{\left( \mathbf{p}\cdot \mathbf{s} \right) \mathbf{p}}{m\left( \omega +m \right)},
\end{eqnarray}
where $\mathbf{s}$ is the spin-direction unit vector in the rest frame of the particle.
By using Eqs. (\ref{eq:PSMDF}) and (\ref{eq:SPO}), we can obtain the spin-resolved single-particle momentum distribution function
\begin{eqnarray}\label{eq:SRSPMDF}
f_{s}^{\pm}&=&\frac{1}{2}f\pm \frac{1}{4\omega}n_0\left( m\mathbbm{a} _0+\mathbf{p}\cdot \mathbf{t}_2 \right) \nonumber\\
&&\mp \frac{1}{4\omega}\mathbf{n}\cdot \left( m\mathbf{a}+\mathbf{p}\times \mathbf{t}_1 \right).
\end{eqnarray}
It can be seen that the spin effect is directly related to $\mathbbm{a}_{\mu}$ and $\mathbbm{t}_{\mu \nu}$ in the DHW functions.

For a spatially homogeneous and time-dependent electric field, $\mathbbm{a} _0$ and $\mathbf{t}_2$ vanish, and Eq. (\ref{eq:SRSPMDF}) becomes
\begin{eqnarray}\label{eq:SPSRSPMDF}
f_{s}^{\pm}=\frac{1}{2}f\mp \frac{1}{4\omega}\mathbf{n}\cdot \left( m\mathbf{a}+\mathbf{p}\times \mathbf{t}_1 \right).
\end{eqnarray}
Furthermore, using one of the transformation relations between the DHW formalism and the QKEs \cite{Aleksandrov2024k},
\begin{eqnarray}\label{eq:CDKL}
\boldsymbol{\lambda }=-\frac{1}{4\omega}\left[ m\mathbf{a}+\mathbf{p}\times \mathbf{t}_1+\frac{\left( \mathbf{p}\cdot \mathbf{a} \right) \mathbf{p}}{\omega +m} \right],
\end{eqnarray}
Eq. (\ref{eq:SPSRSPMDF}) can also be written as
\begin{eqnarray}\label{eq:SPDFL}
f_{s}^{\pm}=\frac{1}{2}f\pm \mathbf{s}\cdot \boldsymbol{\lambda }.
\end{eqnarray}
At $t\rightarrow +\infty$, integrating over full-momentum space with respect to $f_{s}^{\pm}$, one can obtain the spin-resolved particle number density
\begin{eqnarray}\label{eq:SPPND}
N_{s}^{\pm}=\frac{1}{2}N\pm \mathbf{s}\cdot \mathbf{N},
\end{eqnarray}
where $\mathbf{N}=\int{\frac{dq^3}{\left( 2\pi \right) ^3}}\bm{\lambda}\left( \mathbf{q},t \right)$.

Below we consider two special cases of Eq. (\ref{eq:SPSRSPMDF}), where the spin-direction unit vector $\mathbf{s}$ is along the direction of the kinetic momentum $\mathbf{p}$ and the direction of the $z$-axis, respectively.

\subsection{Along the direction of kinetic momentum}\label{subsec:C}
If the spin-direction unit vector $\mathbf{s}$ is parallel to the direction of the kinetic momentum of the particle in the laboratory frame, i.e., $\mathbf{s}=\mathbf{p}/\left|\mathbf{p}\right|$.
The spin polarization vector becomes
\begin{eqnarray}\label{eq:NUhelicity}
n^{\mu}=\left( \frac{\left| \mathbf{p} \right|}{m}, \frac{\omega \mathbf{p}}{m\left| \mathbf{p} \right|} \right).
\end{eqnarray}
According to Eq. (\ref{eq:SRSPMDF}), the spin-resolved momentum distribution function is expressed as
\begin{eqnarray}\label{eq:HRMDF}
f_{h}^{\pm}=\frac{1}{2}f\pm \frac{\left| \mathbf{p} \right|}{4\omega}\left( \mathbbm{a} _0+\frac{\mathbf{p}\cdot \mathbf{t}_2}{m} \right) \mp \frac{\mathbf{p}\cdot \mathbf{a}}{4\left| \mathbf{p} \right|}.
\end{eqnarray}
For spatially homogeneous and time-dependent electric fields, $\mathbbm{a}_0$ and $\mathbf{t}_2$ vanish, and then
\begin{eqnarray}\label{eq:SHHRMDF}
f_{h}^{\pm}=\frac{1}{2}f\mp \frac{\mathbf{p}\cdot \mathbf{a}}{4\left| \mathbf{p} \right|}.
\end{eqnarray}
Based on Eq. (\ref{eq:CDKL}), $\mathbf{p}\cdot\mathbf{a}=-4\mathbf{p}\cdot \boldsymbol{\lambda}$ can be obtained, so in the representation form of QKEs, Eq. (\ref{eq:SHHRMDF}) can also be written as
\begin{eqnarray}\label{eq:SHHRMDFQKE}
f_{h}^{\pm}=\frac{1}{2}f\pm \frac{\mathbf{p}\cdot \boldsymbol{\lambda }}{\left| \mathbf{p} \right|}.
\end{eqnarray}
This equation is just the helicity-resolved momentum distribution function that was defined in Refs. \cite{Aleksandrov2024k,Aleksandrov2024A}.
For example, see Eqs. (14) and (15) in Ref. \cite{Aleksandrov2024A}. Note that positrons are used in our paper.

\subsection{Along the direction of \textit{z}-axis}\label{subsec:D}
Another special form of Eq. (\ref{eq:SPSRSPMDF}) that we consider is the spin-direction unit vector $\mathbf{s}=(0,0,1)$. The spin-resolved momentum distribution function becomes
\begin{eqnarray}\label{eq:ZSRMDF}
f_{s}^{\pm}&=&\frac{1}{2}f\mp \frac{1}{4\omega}\left[ m\mathbf{a}+\mathbf{p}\times \mathbf{t}_1+\frac{\left( \mathbf{p}\cdot \mathbf{a} \right) \mathbf{p}}{\omega +m} \right] _z \nonumber\\
&=&\frac{1}{2}f\pm \lambda _z.
\end{eqnarray}
This is the same as the spin-resolved distribution function Eq. (107) derived by selecting the $z$-axis as the projection direction of spin in Ref. \cite{Aleksandrov2024k}.

In the next section, we will explore the spin-resolved momentum distribution and number density of particles produced by a circularly polarized field for any spin-direction unit vector numerically.

\section{NUMERICAL RESULTS AND ANALYSIS}
\label{sec:four}
The external field we used is the standing wave electric field formed by two counter-propagating laser beams along the $z$-axis:
\begin{eqnarray}\label{eq:REF}
\mathbf{E}\left( t \right) =\frac{E_0}{\sqrt{2}}e^{-\frac{\left( \Omega t \right) ^2}{\sigma ^2}}\left( \begin{array}{c}
	\cos \left( \Omega t \right)\\
	\sin \left( \Omega t \right)\\
	0\\
\end{array} \right),
\end{eqnarray}
where $E_0/\sqrt{2}$ and $\Omega$ are the amplitude and frequency of the electric field, respectively, and the dimensionless parameter $\sigma$ controls the duration of the electric field pulse.
In the numerical calculations, we set the initial time $t_{\mathrm{in}}$ and the final time $t_{\mathrm{out}}$ to a sufficiently large value to ensure that the external electric field $\mathbf{E}(t)$ asymptotically becomes $0$ and the results of the calculations converge.
In addition, we set the initial value of the vector potential to $\mathbf{A}(t_{\mathrm{in}})=0$.

Before specific numerical calculations, we theoretically analyze the spin-resolved momentum distribution function in the polarization plane $(q_x, q_y)$ at $q_z=0$ with the assistance of QKEs.
In this case, the momentum direction of created particles is perpendicular to the $z$-axis, i.e., $\mathbf{q}=\mathbf{q}_{\bot}$, where $\bot$ represents perpendicular to the $z$-axis.
Since the electric field (\ref{eq:REF}) we used is also perpendicular to the $z$-axis, the quantities $\boldsymbol{\xi }_1$ and $\boldsymbol{\xi }_2$ in Eq. (\ref{eq:QKESXI}) can be written as $\boldsymbol{\xi }_1=\xi _1\mathbf{e}_z$ and $\boldsymbol{\xi }_2=\xi _2\mathbf{e}_{\bot}$,
where $\mathbf{e}_z$ and $\mathbf{e}_{\bot}$ are unit vectors along the $z$-axis and perpendicular to the $z$-axis, respectively.
By decomposing the vector functions $\boldsymbol{\lambda}$, $\boldsymbol{\mu}$, and $\boldsymbol{\nu}$ into $\boldsymbol{\lambda}=\lambda_z\mathbf{e}_z+\boldsymbol{\lambda }_{\bot}$, $\boldsymbol{\mu}=\mu_z\mathbf{e}_z+\boldsymbol{\mu }_{\bot}$, and $\boldsymbol{\nu}=\nu_z\mathbf{e}_z+\boldsymbol{\nu }_{\bot}$, the system of QKEs shown in Eq. (\ref{eq:QKES}) can be transformed into the following form:
\begin{eqnarray}\label{eq:XYQKES}
\begin{array}{l}
\dot{f}=\displaystyle{-4\xi _2\mathbf{e}_{\bot}\cdot \boldsymbol{\mu }_{\bot}},
\\
\\
\dot{\lambda}_z=\displaystyle{-2\xi _2\mathbf{e}_z\cdot \left( \mathbf{e}_{\bot}\times \boldsymbol{\nu }_{\bot} \right)},
\\
\\
\dot{\boldsymbol{\lambda}}_{\bot}=\displaystyle{2\xi _1\mathbf{e}_z\times \boldsymbol{\lambda }_{\bot}-2\xi _2\nu _z\mathbf{e}_{\bot}\times \mathbf{e}_z},
\\
\\
\dot{\mu}_z=\displaystyle{2\omega \nu _z},
\\
\\
\dot{\boldsymbol{\mu}}_{\bot}=\displaystyle{2\xi _1\mathbf{e}_z\times \boldsymbol{\mu }_{\bot}+\xi _2\left( f-1 \right) \mathbf{e}_{\bot}+2\omega \boldsymbol{\nu }_{\bot}},
\\
\\
\dot{\nu}_z=\displaystyle{-2\xi _2\mathbf{e}_z\cdot \left( \mathbf{e}_{\bot}\times \boldsymbol{\lambda }_{\bot} \right) -2\omega \mu _z},
\\
\\
\dot{\boldsymbol{\nu}}_{\bot}=\displaystyle{2\xi _1\mathbf{e}_z\times \boldsymbol{\nu }_{\bot}-2\xi _2\lambda _z\mathbf{e}_{\bot}\times \mathbf{e}_z-2\omega \boldsymbol{\mu }_{\bot}}.
\end{array}
\end{eqnarray}
Notice that $\dot{\boldsymbol{\lambda}}_{\bot}\cdot \boldsymbol{\lambda }_{\bot}+\dot{\nu}_z\nu _z+\dot{\mu}_z\mu _z=0$.
According to the initial conditions of the QKEs, we get $\boldsymbol{\lambda }_{\bot}^{2}+\mu _{z}^{2}+\nu _{z}^{2}=0$, so $\lambda _x=\lambda _y=\mu _z=\nu _z=0$.
Then the spin-resolved momentum distribution function given in Eq. (\ref{eq:SPDFL}) becomes
\begin{eqnarray}\label{eq:XYSPDFL}
f_{s}^{\pm}=\frac{1}{2}f\pm s_z\lambda _z.
\end{eqnarray}
This means that the spin-resolved momentum distribution in the polarization plane at $q_z=0$ is only related to $s_z$ and has nothing to do with $s_x$ and $s_y$.

\subsection{Spin-resolved momentum distributions}\label{subsec:E}
In this section, we discuss the spin-resolved momentum distribution function in two cases.
The first one is $s_z=0$ and the second one is $s_z\ne 0$.

\begin{figure*}[!ht]
\centering
\includegraphics[width=0.98\textwidth]{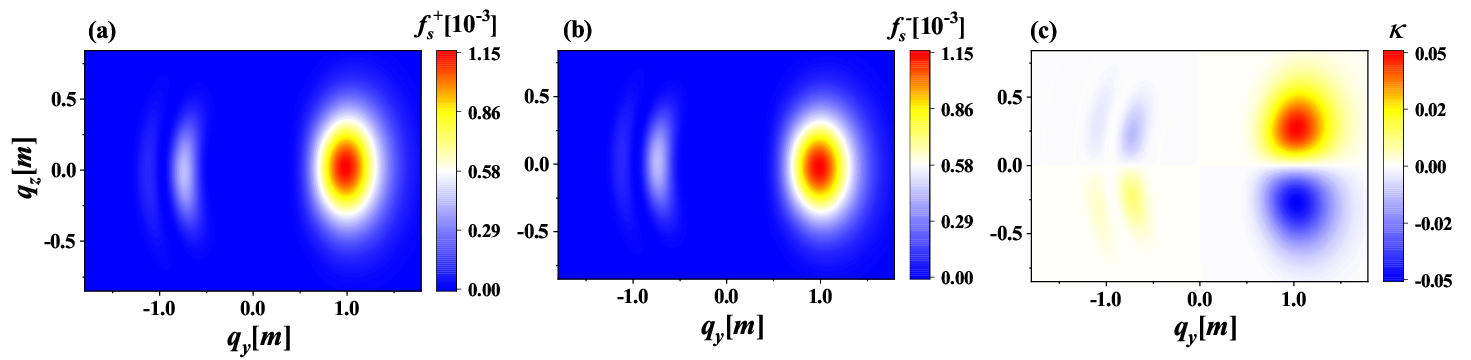}
\includegraphics[width=0.98\textwidth]{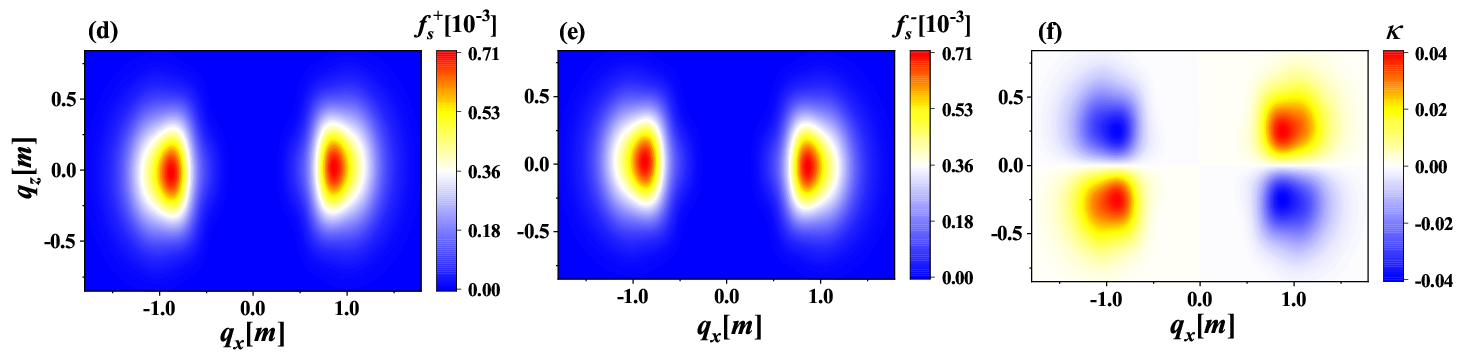}
\caption{The spin-resolved momentum distribution functions and the spin asymmetry degree $\kappa$ of positrons for selecting $\mathbf{s}=(1/\sqrt{2}, 1/\sqrt{2}, 0)$. Panels (a) and (d) correspond to the spin-up case, while panels (b) and (e) correspond to the spin-down case. Panels (c) and (f) denote the spin asymmetry degree.
The first and second rows correspond to the momentum planes $(q_x=0, q_y, q_z)$ and $(q_x, q_y=0, q_z)$, respectively.
The electric field parameters are $E_0=0.5E_{\mathrm{cr}}$, $\Omega=0.5m$, and $\sigma=5$.
\label{fig:110}}
\end{figure*}

First, we choose the spin-direction unit vector as $\mathbf{s}=(1/\sqrt{2}, 1/\sqrt{2}, 0)$.
According to Eq. (\ref{eq:XYSPDFL}), the momentum distributions for both spin-up and spin-down particles are equal to $f/2$ in the polarization plane at $q_z=0$.
There is no spin asymmetry.
Thus, we only focus on the distribution functions $f_{s}^{\pm}$ in the momentum planes $(q_x=0, q_y, q_z)$ and $(q_x, q_y=0, q_z)$.
The spin-resolved momentum distributions and spin asymmetry of positrons produced by the external rotating field (\ref{eq:REF}) are exhibited in Fig. \ref{fig:110}.
Here the spin asymmetry degree $\kappa$ is defined as $f_{s}^{+}-f_{s}^{-}$ divided by the maximum value of the total momentum distribution function $f$.

As can be seen from Fig. \ref{fig:110}(a) and (b), the patterns of the spin-up distribution function $f_{s}^{+}$ and the spin-down distribution function $f_{s}^{-}$ are similar.
The maximum values of both $f_{s}^{+}$ and $f_{s}^{-}$ are about $f/2$.
Although the positions of the maxima have only a small shift (about $0.02 m$) in the positive $z$-axis direction and the negative $z$-axis direction, respectively, there is still a maximum of about $4.6\%$ spin asymmetry at $(q_y, q_z)\approx(1.03m, \pm0.27m)$, see Fig. \ref{fig:110}(c).
Furthermore, the spin asymmetry degree is shifted significantly in the positive and negative $z$-axis directions, respectively.
By analyzing the spin asymmetry degree $\kappa$ and the momentum distribution function $f_{s}^{\pm}$, the following phenomena are observed:
In the right region of the distribution function ($q_y>0$), the spin-up (spin-down) positrons prefer to move towards the positive $z$-axis (negative $z$-axis) direction. In the left region of the distribution function ($q_y<0$), the spin-up (spin-down) positrons prefer to move along the negative $z$-axis (positive $z$-axis) direction.
This is different from the result in Ref. \cite{Aleksandrov2024A}, where spin-up (or spin-down) positrons always prefer to move along the positive $z$-axis (or negative $z$-axis) direction.
Moreover, both $f_{s}^{+}$ and $f_{s}^{-}$ are asymmetric with respect to $q_z=0$, as can be found in Eq. (\ref{eq:SPDFL}).
This is because the total momentum distribution function $f$ is symmetric about $q_z=0$.
However, according to the QKEs shown in Eq. (\ref{eq:QKES}), for the reflection transformation $q_z\rightarrow -q_z$, the sign of $\lambda _z$ remains the same while $\lambda _x$ and $\lambda _y$ change sign \cite{Aleksandrov2024A}.
It also explains why the spin asymmetry degree $\kappa$ is an odd function of $q_z$.
In addition, in Figs. \ref{fig:110}(a) and (b), the maximum value of $f_{s}^{\pm}$ is significantly shifted along the positive $y$-axis direction.
This characteristic is the same as the total momentum distribution function.
Since the time integral of $E_y(t)$ over the interval $[0, t_{\mathrm{out}}]$ is a sufficiently large positive value, positrons (or electrons) produced mainly near $t=0$ acquire a net positive (or negative) momentum due to the influence of this field.

For the momentum plane $(q_x, q_y=0, q_z)$, the maximum value of the absolute value of the spin asymmetry degree is approximately $3.9\%$, see Fig. \ref{fig:110}(f). Its position has a shift of about $0.26m$ along the $z$-axis direction.
In the region where $q_x>0$, the spin-up positrons prefer to move toward the positive $z$-axis direction, while in the region $q_x<0$, they prefer to move along the negative $z$-axis direction.
Positrons with opposite spin directions exhibit reversed directional preferences in their motion.
The symmetry of the momentum distribution can be analyzed through QKEs shown in Eq. (\ref{eq:QKES}).
For the electric field (\ref{eq:REF}), the signs of $f(q_{x}^{*}, q_y, q_z, t)$, $\lambda_y(q_{x}^{*}, q_y, q_z, t)$, and $\lambda_z(q_{x}^{*}, q_y, q_z, t)$ are unchanged under the time reversal $t\rightarrow -t$ and the spatial reflection transformation $q_{x}^{*}\rightarrow -q_{x}^{*}$, but $\lambda_x(q_{x}^{*}, q_y, q_z, t)$ changes sign.
Here $q_{x}^{*}=q_x-A_x(t_{\mathrm{out}})/2$.
Therefore, the single-particle momentum distribution function $f$ is symmetric about $q_{x}^{*}$, but $f_{s}^{\pm}$ is not symmetric about $q_{x}^{*}$.
Moreover, based on these symmetry relations, it can also be found that $\kappa$ is not anti-symmetric with respect to the $q_x$-axis.
The reason why $\kappa$ appears to be anti-symmetric in Fig. \ref{fig:110}(f) is that the values of $\lambda_y$ are much smaller than those of $\lambda_x$ in the momentum plane $(q_x, q_y=0, q_z)$.

\begin{figure*}[t]
\centering
\includegraphics[width=0.98\textwidth]{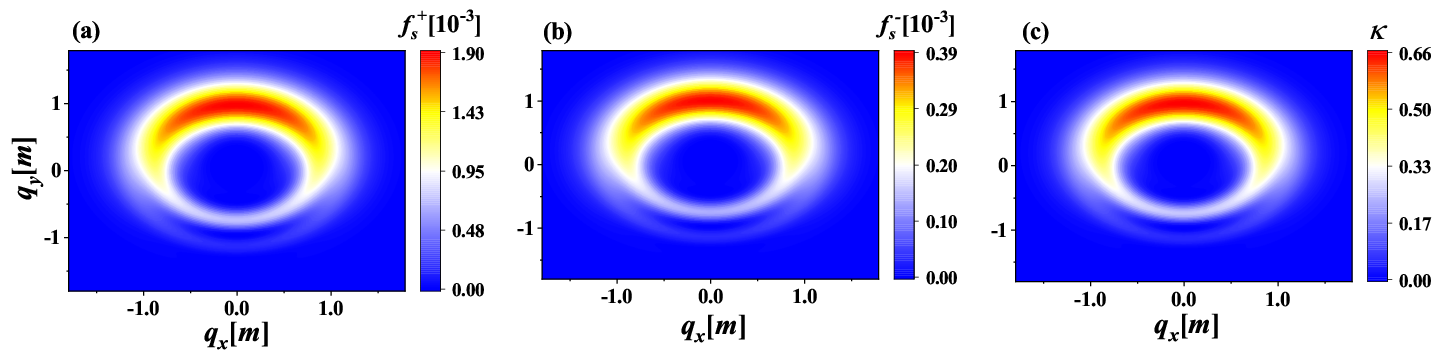}
\includegraphics[width=0.98\textwidth]{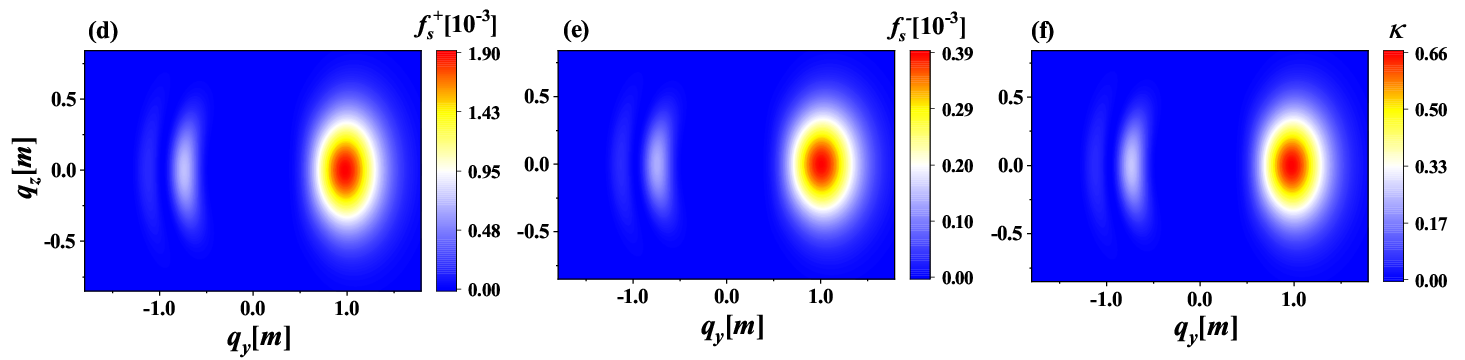}
\includegraphics[width=0.98\textwidth]{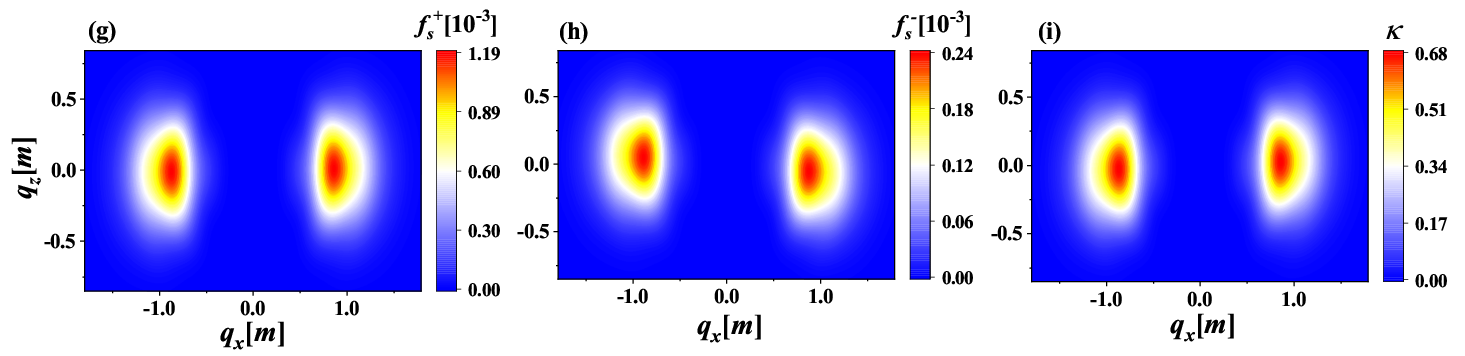}
\caption{The spin-resolved momentum distribution functions and the spin asymmetry degree $\kappa$ of positrons for choosing $\mathbf{s}=(1/\sqrt{2}, 0, 1/\sqrt{2})$. Panels (a), (d), and (g) correspond to the spin-up case, while (b), (e), and (h) represent the spin-down case. Panels (c), (f), and (i) denote the spin asymmetry degree.
The first, second, and third rows correspond to the momentum planes $(q_x, q_y, q_z=0)$, $(q_x=0, q_y, q_z)$, and $(q_x, q_y=0, q_z)$, respectively.
The electric field parameters are $E_0=0.5E_{\mathrm{cr}}$, $\Omega=0.5m$, and $\sigma=5$.
\label{fig:101}}
\end{figure*}

In addition, we have also calculated the spin asymmetry degree for the case where $s_z=0$, $s_x$ and $s_y$ take other values.
The results show that the maximum value of the absolute value of the spin asymmetry is usually within $6.5\%$ in the momentum planes $(q_x=0, q_y, q_z)$ and $(q_x, q_y=0, q_z)$.

By choosing $\mathbf{s}=(1/\sqrt{2}, 0, 1/\sqrt{2})$, the spin-resolved momentum distribution functions and spin asymmetry degrees in the momentum planes $(q_x, q_y, q_z=0)$, $(q_x=0, q_y, q_z)$, and $(q_x, q_y=0, q_z)$ are calculated and shown in Fig. \ref{fig:101}.

The characteristics of the total momentum distribution in the momentum plane $(q_x, q_y, q_z=0)$ have been studied previously, e.g., Refs. \cite{Blinne2014p,Li2015E}.
As can be seen in Figs. \ref{fig:101}(a) and (b), the characteristics of the spin-up and spin-down momentum distributions $f_{s}^{\pm}$ are similar to those of the total momentum distribution.
One of the most obvious characteristics is that the momentum distribution is mainly concentrated in the momentum plane $q_y>0$.  The reason is the same as our previous discussion, i.e., influenced by the electric field component $E_y(t)$.
Of course, it can also be explained by analyzing the distribution of turning points in complex time plane \cite{Dunne2011,Dumlu2010t}.
Another obvious characteristic in Figs. \ref{fig:101}(a) and (b) is the interference patterns of the spin-resolved momentum distributions in the negative $y$-axis direction.
The reason for the appearance of interference patterns is still attributed to the effect of $E_y(t)$.
Therefore, the difference between $f_{s}^{+}$ and $f_{s}^{-}$ lies mainly in their numerical values, with $f_{s}^{+}$ being about $4.87$ times that of $f_{s}^{-}$.
However, in Ref. \cite{Hu2024}, the spin-up and spin-down momentum distribution functions differ by two orders of magnitude.
This is because the difference between the values of $f_{s}^{+}$ and $f_{s}^{-}$ is maximized at $s_z=\pm1$ as can be seen from Eq. (\ref{eq:XYSPDFL}).
In addition, if the direction of the unit vector $\mathbf{s}$ is reversed, i.e., having the form of $\mathbf{s}=(-1/\sqrt{2}, 0, -1/\sqrt{2})$, we can obtain $f_{s,\mathbf{s}}^{\pm}=f_{s,-\mathbf{s}}^{\mp}$ from Eq. (\ref{eq:SPSRSPMDF}).
This is simply the result of changing the observation direction of the spin.
Interestingly, according to Eq. (\ref{eq:XYSPDFL}), this symmetry relation also holds if only $s_z$ changes sign, i.e., $f_{s,s_z}^{\pm}=f_{s,-s_z}^{\mp}$.

In Figs. \ref{fig:101}(d) and (e), one can see that $f_{s}^{+}$ and $f_{s}^{-}$ in the momentum plane $(q_x=0, q_y, q_z)$ have the same shape and are both approximately symmetric about $q_z=0$. The value of $f_{s}^{+}$ is about $4.87$ times that of $f_{s}^{-}$.
However, as observed in Fig. \ref{fig:101}(h), spin-down positrons exhibit a pronounced directional preference in their motion within the momentum plane $(q_x, q_y=0, q_z)$.
For the spin asymmetry degree, the maximum values of $\kappa$ in Figs. \ref{fig:101}(c), (f), and (i) can reach about $66\%$, $66\%$, and $68\%$, respectively.
This is also an important characteristic of spin asymmetry when the spin-direction unit vector $\mathbf{s}$ has a large $z$-component.

\subsection{Particle number density}\label{subsec:F}
\begin{figure}[!ht]
\centering
\includegraphics[width=0.48\textwidth]{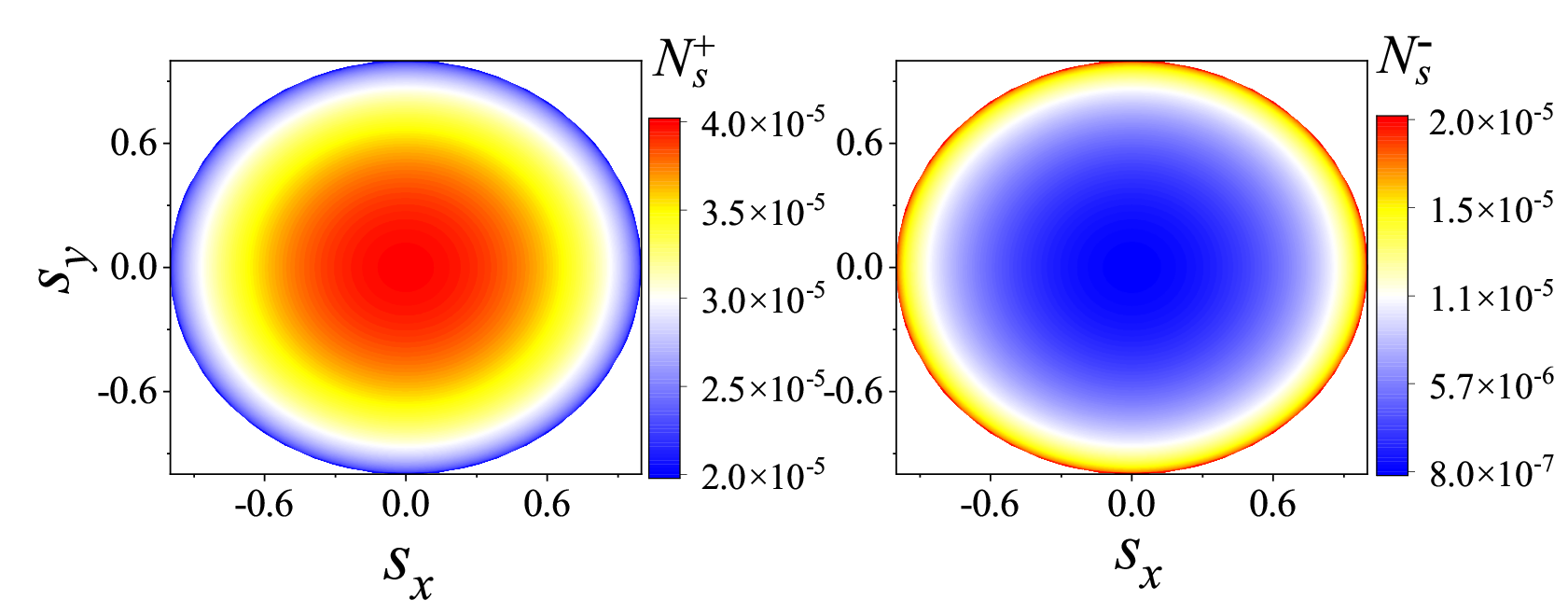}%
\caption{The spin-up particle number density (left panel) and the spin-down particle number density (right panel) for any spin-direction unit vector $\mathbf{s}=\left( s_x, s_y, \sqrt{1-s_{x}^{2}-s_{y}^{2}} \right)$.
The electric field parameters are $E_0=0.3\sqrt{2}E_{\mathrm{cr}}$, $\Omega=0.65m$, and $\sigma=13\sqrt{2}$.
\label{fig:nupdown}}
\end{figure}

According to Eq. (\ref{eq:SPPND}), we can calculate the spin-resolved particle number density $N_{s}^{\pm}$ for any spin direction unit vector, as shown in Fig. \ref{fig:nupdown}.
Since $N_{s}^{\pm}$ is antisymmetric with respect to $\mathbf{s}$, only the top view of the upper half of the sphere (i.e., $s_z=\sqrt{1-s_{x}^{2}-s_{y}^{2}}$ ) is shown here.
One can see that $N_{s}^{+}$ (or $N_{s}^{-}$) reaches its maximum (or minimum) value at $s_z=1$ and decreases (or increases) as $s_z$ decreases.
At $s_z=0$, $N_{s}^{+}$ equals $N_{s}^{-}$.
This is because when $E_z\left( t \right) \equiv 0$, $\lambda _x\left( q_z \right) =-\lambda _x\left( -q_z \right) $ and $\lambda _y\left( q_z \right) =-\lambda _y\left( -q_z \right) $, resulting in $N_{s}^{\pm}=N/2\pm s_zN_z=N/2$.
For a general three-component electric field, the maximum or minimum value of the spin-resolved particle number density is obtained at $\mathbf{s}_N=\mathbf{N} /\left| \mathbf{N} \right|$, see Eq. (\ref{eq:SPPND}).
The physical phenomenon that we are more concerned with is the spin asymmetry.
Therefore, we define the spin asymmetry degree
\begin{eqnarray}\label{eq:NSAD}
\kappa _N=\frac{N_{s}^{+}-N_{s}^{-}}{N_{s}^{+}+N_{s}^{-}}.
\end{eqnarray}
Note that the spin up and spin down are relative to the unit vector $\mathbf{s}_N$.
In the following, the effects of the electric field strength and field frequency on $\kappa_N$ are investigated.

\begin{figure}[!ht]
\centering
\includegraphics[width=0.45\textwidth]{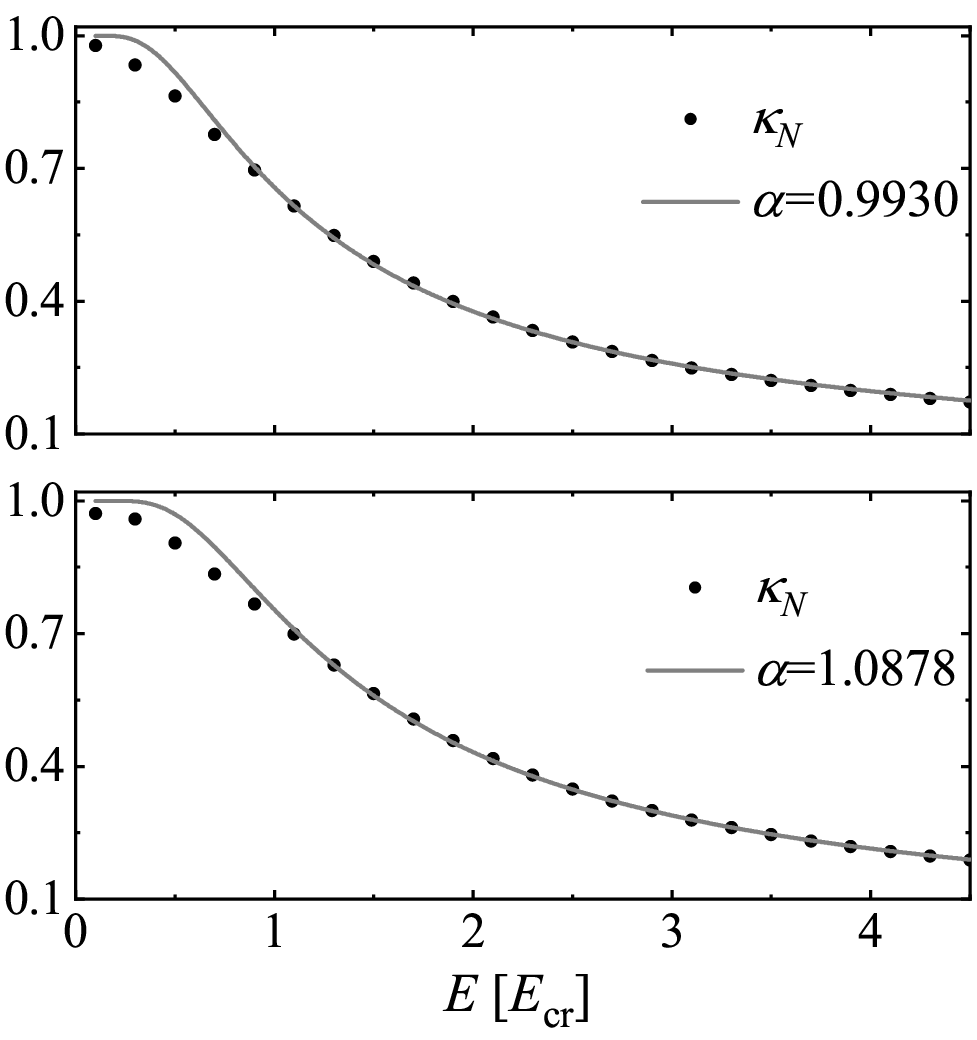}%
\caption{The spin asymmetry degree $\kappa_N$ as a function of the electric field amplitude $E=E_0/\sqrt{2}$ for the spin-direction unit vector $\mathbf{s}=(0, 0, 1)$.
The solid gray line represents the fitted curve for $\kappa_N$.
Other electric field parameters are $\sigma=10\sqrt{2}$ and $\Omega=0.5m$ for the upper panel, and $\sigma=13\sqrt{2}$ and $\Omega=0.65m$ for the lower panel.
\label{fig:nefit}}
\end{figure}

The relationship between $\kappa_N$ and the electric field amplitude $E$ for the spin-direction unit vector $\mathbf{s}=(0, 0, 1)$ is shown in Fig. \ref{fig:nefit}.
The results demonstrate that when the electric field amplitude $E=0.1E_{\mathrm{cr}}$ and $\zeta=5$ (where $\zeta=m\omega/(eE)$ is the Keldysh adiabatic parameter), the number density of spin-up particles is approximately $90$ times that of spin-down particles.
One can also see that in the transition region from perturbative to non-perturbative processes where $\zeta \sim O\left( 1 \right)$, the spin asymmetry degree $\kappa_N$ decreases rapidly with the electric field amplitude $E$.
When $\zeta \ll 1$, the variation of $\kappa_N$ with $E$ becomes gradual.
The contribution of spin-down particles to the total particle number density becomes important.
For example, when $E=2.5E_{\mathrm{cr}}$, the number density of spin-up particles is only about $2$ times that of spin-down particles.
The reason can be explained as follows.
We know that there are two primary vacuum pair production mechanisms: the Schwinger tunneling mechanism (when $\zeta \ll 1$) and the multiphoton absorption mechanism (when $\zeta \gg 1$).
For the left-handed circularly polarized electric field (\ref{eq:REF}), photons have well-defined spin angular momentum with a projection along the $z$-axis of $+1$.
For a small field strength, pair production is dominated by multiphoton absorption.
The spin angular momentum of absorbed photons is converted into the angular momentum of created electron-positron pairs, which prefers to induce the production of particles with spin $+1/2$ along the $z$-axis.
However, with the increase of the field strength, pair production becomes dominated by the Schwinger tunneling mechanism.
For tunneling pair production, the angular momentum of the external field mainly transfers to the orbital angular momentum of created pairs due to the more pronounced acceleration of created particles by the strong field. Consequently, the dependence of tunneling probability on particle spin states is relatively weak. For an extremely high field strength, the number density of produced spin-up and spin-down particles is nearly comparable.
Moreover, we fit the relationship between the spin asymmetry degree and the electric field strength, and obtain
\begin{eqnarray}\label{eq:NSADFIT}
\kappa _N=\frac{1-\exp \left( -\pi \zeta ^{\alpha} \right)}{1+\exp \left( -\pi \zeta ^{\alpha} \right)},
\end{eqnarray}
where $\alpha$ is a fitting parameter, see the gray solid line in Fig. \ref{fig:nefit}.
For the upper panel of Fig. \ref{fig:nefit}, $\alpha=0.9930$. The goodness of fit $\mathrm{R}^2$ and $95\%$ confidence interval are $0.9948$ and $\left[0.9720, 1.0140 \right]$, respectively.
For the lower panel of Fig. \ref{fig:nefit}, $\alpha=1.0878$. The goodness of fit $\mathrm{R}^2$ and $95\%$ confidence interval are $0.9924$ and $\left[1.0587, 1.1169 \right]$, respectively.
It can be seen that in the region dominated by tunneling pair production, the fitting results and numerical results show excellent agreement.

\begin{figure}[!ht]
\centering
\includegraphics[width=0.48\textwidth]{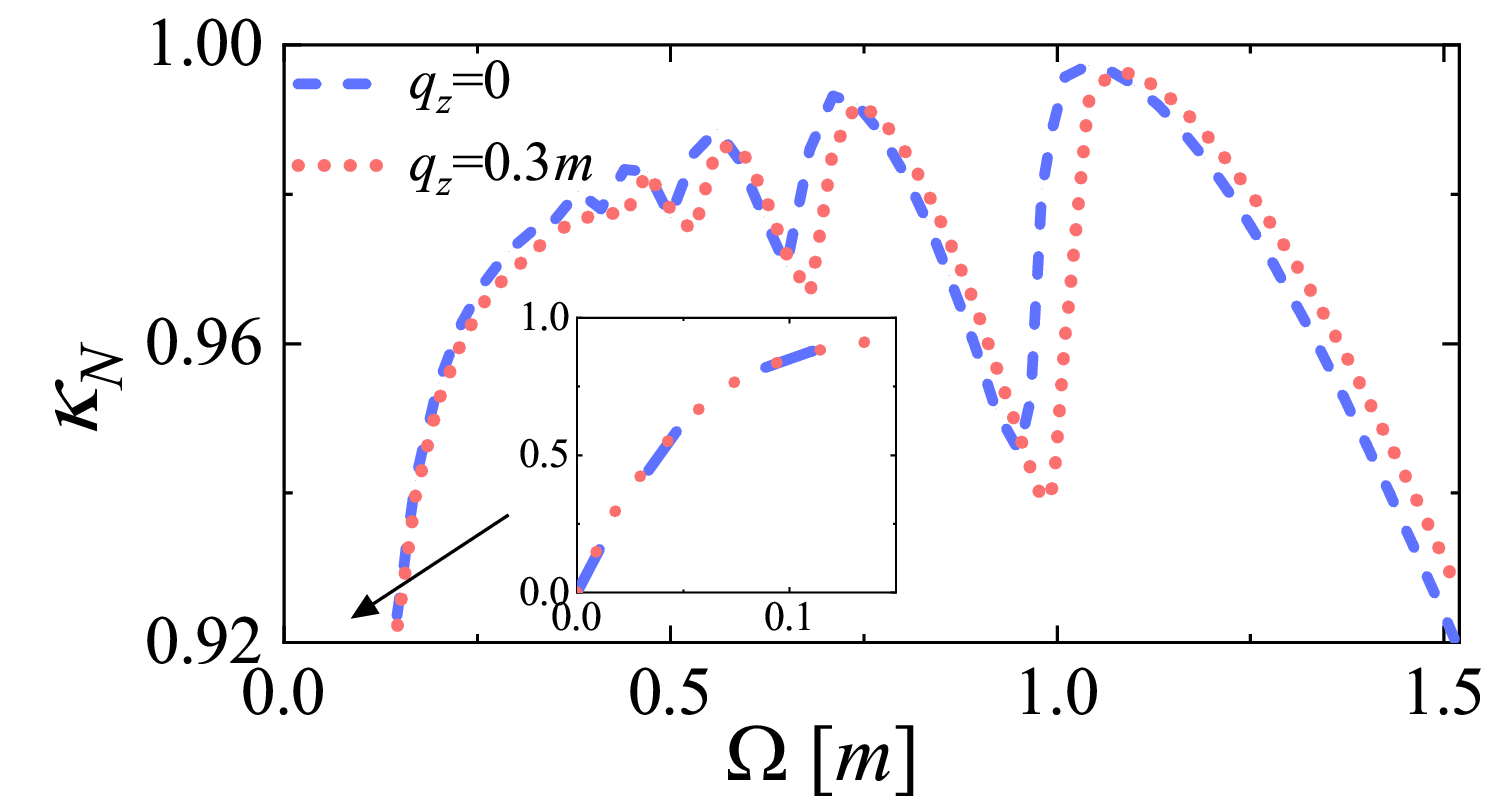}%
\caption{The spin asymmetry degree $\kappa_N$ as a function of the electric field frequency $\Omega$ for the spin-direction unit vector $\mathbf{s}=(0, 0, 1)$.
The blue dashed line and red dotted line represent the results in the momentum planes $q_z=0$ and $q_z=0.3m$, respectively.
Other electric field parameters are $E_0=0.1\sqrt{2}E_{\mathrm{cr}}$ and $\sigma=20\sqrt{2}\Omega/m$.
\label{fig:knw}}
\end{figure}

The relationship between $\kappa_N$ and the electric field frequency $\Omega$ for the spin-direction unit vector $\mathbf{s}=(0, 0, 1)$ is shown in Fig. \ref{fig:knw}.
Note that the spin asymmetry degree is computed in two typical momentum planes, $q_z=0$ and $q_z=0.3m$, rather than the full momentum space.
As shown in Fig. \ref{fig:knw}, in the low-frequency region ($\Omega<0.4m$ and $\zeta<4$), $\kappa_N$ increases with the increase of $\Omega$.
For linearly polarized fields (i.e., $\Omega=0$ in Eq. (\ref{eq:REF})), the average spin angular momentum of photons is $0$.
Therefore, there is no net spin angular momentum transfer. The number density of spin-up particles is exactly equal to that of spin-down particles.
Figure \ref{fig:knw} also shows that for subcritical electric fields, the weakening of multiphoton absorption can decrease the spin asymmetry degree, providing another perspective to verify the conclusion drawn in Fig. \ref{fig:nefit}.
In the high-frequency region ($\Omega>0.5m$ and $\zeta>5$), $\kappa_N$ has some local maximum in the vicinity of the threshold for multiphoton absorption. These local maximum values increase as $\Omega$ increases.
This is because particle pair production is most efficient at the threshold of multiphoton absorption, where the transfer of angular momentum is also more significant.
Moreover, we find that the results for $q_z=0$ and $q_z=0.3m$ exhibit similar variation trends.
According to the formula of the total energy of particles, it is easy to see that $q_z$ plays a role similar to the particle mass, which will affect the frequency threshold of multiphoton absorption.
Finally, we can predict that in the full momentum space $\kappa_N$ and $\Omega$ have a similar relationship to those shown in Fig. \ref{fig:knw}.

\section{CONCLUSIONS and discussions}
\label{sec:five}
In this study, a general spin-resolved momentum distribution function of electron-positron pairs produced from the vacuum under any external field was derived by combining the covariant spin projection operator and the Dirac-Heisenberg-Wigner (DHW) formalism.
It is found that the helicity-resolved momentum distribution function is actually a special case of the spin-resolved momentum distribution function.
It can be obtained by choosing the spin-direction unit vector along the momentum direction of produced particles.
In particular, for a spatially homogeneous and time-dependent electric field, the helicity-resolved momentum distribution we derived is equivalent to the one obtained from the quantum kinetic equations (QKEs) in \cite{Aleksandrov2024k,Aleksandrov2024A}.

The spin-resolved momentum distribution and number density of particles produced by a left-handed circularly polarized electric field for any spin-direction unit vector are also investigated numerically.
The results show that when the $z$-component of the spin-direction unit vector is zero, the number density of produced particles with spin up and spin down is equal, but their momentum distributions exhibit asymmetry.
The particles with specific spin polarization have directional preferences in their motion.
In the positive (or negative) momentum direction $q_x$ or $q_y$, spin-up positrons preferentially move toward the positive $z$-axis (or negative $z$-axis), while spin-down positrons show opposite directional preferences during their motion.
Moreover, the $z$-component of the spin-direction unit vector $s_z$ also has significant effect on spin asymmetry.
There may even be a difference of $1-3$ orders of magnitude between the spin-up and spin-down particle number density when $s_z$ is nonzero.

By examining the effects of the electric field strength and frequency on the spin-resolved particle number density, we have discovered the following results.
The difference of $1-3$ orders of magnitude between the spin-up and spin-down particle number density is caused by angular momentum transfer during the multiphoton absorption process.
For left-handed (or right-handed) circularly polarized fields, both electrons and positrons produced primarily have spin values of $+1/2$ (or $-1/2$) along the propagation direction of the field.
As the electric field strength increases and/or the field frequency decreases, the difference between the spin-up and spin-down particle number density decreases rapidly during the transition phase from perturbation to nonperturbation regime.
This is because the influence of particle spin states on tunneling pair production is relatively weak.
Interestingly, these results also suggest that spin asymmetry degree may serve as a tool for determining whether multiphoton absorption or tunneling dominates.

The study of spin-resolved momentum distributions has deepened our understanding of how particle pairs are produced from the vacuum for fields carrying angular momentum.
Since the spin-resolved momentum distribution we derived also applies to spatially inhomogeneous fields, it provides the possibility for future studies on how angular momentum is transferred from any external field to matter under extreme conditions.

\begin{acknowledgments}
The work is supported by the National Natural Science Foundation of China (NSFC) under Grants No. 11974419 and No. 11705278, the Strategic Priority Research Program of Chinese Academy of Sciences (Grant No. XDA25051000, XDA25010100), and the Fundamental Research Funds for the Central Universities (No. 2023ZKPYL02, 2025 Basic Sciences Initiative in Mathematics and Physics).
\end{acknowledgments}

\end{document}